\documentclass[preprint]{elsarticle}

\usepackage{hyperref}

\usepackage{xspace}

\usepackage{upgreek}

\journal{Physics Letters B}

\begin{document}
\newcommand{\ind}[2]{\ensuremath{#1_{\mathrm{#2}}}}
\newcommand{\dind}[3]{\ensuremath{#1_{\mathrm{#2}}^{\mathrm{#3}}}}
\renewcommand{\deg}[1]{\ensuremath{#1^{\circ}}\xspace}
\begin{frontmatter}

\title{A comparison of the cosmic-ray energy scales of Tunka-133 and KASCADE-Grande via their radio extensions Tunka-Rex and LOPES}

\author[a]{W.D.~Apel}
\author[b]{J.C.~Arteaga-Vel\'azquez}
\author[c]{L.~B\"ahren}
\author[d]{P.A.~Bezyazeekov}
\author[a]{K.~Bekk}
\author[e]{M.~Bertaina}
\author[f]{P.L.~Biermann}
\author[a,g]{J.~Bl\"umer}
\author[a]{H.~Bozdog}
\author[h]{I.M.~Brancus}
\author[d]{N.M.~Budnev}
\author[e,i]{E.~Cantoni}
\author[e]{A.~Chiavassa}
\author[a]{K.~Daumiller}
\author[j]{V.~de~Souza}
\author[e]{F.~Di~Pierro}
\author[a]{P.~Doll}
\author[a]{R.~Engel}
\author[c,f,k]{H.~Falcke}
\author[d]{O.~Fedorov}
\author[g]{B.~Fuchs}
\author[l]{H.~Gemmeke}
\author[d]{O.A.~Gress}
\author[m]{C.~Grupen}
\author[a]{A.~Haungs}
\author[a]{D.~Heck}
\author[a]{R.~Hiller\corref{mycorrespondingauthor}}
\author[k]{J.R.~H\"orandel}
\author[f]{A.~Horneffer}
\author[g]{D.~Huber}
\author[a]{T.~Huege}
\author[n]{P.G.~Isar}
\author[o]{K.-H.~Kampert}
\author[g]{D.~Kang}
\author[d]{Y.~Kazarina}
\author[l]{M.~Kleifges}
\author[p]{E.E.~Korosteleva}
\author[a]{D.~Kostunin}
\author[l]{O.~Kr\"omer}
\author[k]{J.~Kuijpers}
\author[p]{L.A.~Kuzmichev}
\author[g]{K.~Link}
\author[p]{N.~Lubsandorzhiev}
\author[q]{P.~{\L}uczak}
\author[g]{M.~Ludwig}
\author[a]{H.J.~Mathes}
\author[g]{M.~Melissas}
\author[d]{R.R.~Mirgazov}
\author[d]{R.~Monkhoev}
\author[i]{C.~Morello}
\author[a]{J.~Oehlschl\"ager}
\author[p]{E.A.~Osipova}
\author[d]{A.~Pakhorukov}
\author[g]{N.~Palmieri}
\author[d]{L.~Pankov}
\author[a]{T.~Pierog}
\author[p]{V.V.~Prosin}
\author[o]{J.~Rautenberg}
\author[a]{H.~Rebel}
\author[a]{M.~Roth}
\author[r]{G.I.~Rubtsov}
\author[l]{C.~R\"uhle}
\author[h]{A.~Saftoiu}
\author[a]{H.~Schieler}
\author[l]{A.~Schmidt}
\author[a]{S.~Schoo}
\author[a]{F.G.~Schr\"oder}
\author[s]{O.~Sima}
\author[h]{G.~Toma}
\author[i]{G.C.~Trinchero}
\author[a]{A.~Weindl}
\author[t]{R.~Wischnewski}
\author[a]{J.~Wochele}
\author[q]{J.~Zabierowski}
\author[d]{A.~Zagorodnikov}
\author[f]{J.A.~Zensus}
\author{(Tunka-Rex and LOPES Collaborations)}

\address[a]{Institut f\"ur Kernphysik, Karlsruhe Institute of Technology (KIT), Karlsruhe, Germany}
\address[b]{Instituto de F\'isica y Matem\'aticas, Universidad Michoacana, Morelia, Michoac\'an, Mexico}
\address[c]{ASTRON, Dwingeloo, The Netherlands}
\address[d]{Institute of Applied Physics ISU, Irkutsk, Russia}
\address[e]{Dipartimento di Fisica, Universit\`a degli Studi di Torino, Torino, Italy}
\address[f]{Max-Planck-Institut f\"ur Radioastronomie, Bonn, Germany}
\address[g]{Institut f\"ur Experimentelle Kernphysik, Karlsruhe Institute of Technology (KIT), Karlsruhe, Germany}
\address[h]{National Institute of Physics and Nuclear Engineering, Bucharest-Magurele, Romania}
\address[i]{Osservatorio Astrofisico di Torino, INAF Torino, Torino, Italy}
\address[j]{Instituto de F\'isica de S\~ao Carlos, Universidade de S\~ao Paulo, S\~ao Carlos, Brasil}
\address[k]{Department of Astrophysics, Radboud University Nijmegen, AJ Nijmegen, The Netherlands}
\address[l]{Institut f\"ur Prozessdatenverarbeitung und Elektronik, Karlsruhe Institute of Technology (KIT), Karlsruhe, Germany}
\address[m]{Faculty of Natural Sciences and Engineering, Universit\"at Siegen, Siegen, Germany}
\address[n]{Institute for Space Sciences, Bucharest-Magurele, Romania}
\address[o]{Fachbereich C, Physik, Bergische Universit\"at Wuppertal, Wuppertal, Germany}
\address[p]{Skobeltsyn Institute of Nuclear Physics MSU, Moscow, Russia}
\address[q]{Department of Astrophysics, National Centre for Nuclear Research, {\L}\'{o}d\'{z}, Poland}
\address[r]{Institute for Nuclear Research of the Russian Academy of Sciences, Moscow, Russia}
\address[s]{Department of Physics, University of Bucharest, Bucharest, Romania}
\address[t]{DESY, Zeuthen, Germany}


\cortext[mycorrespondingauthor]{Corresponding author: roman.hiller@kit.edu}



\begin{abstract}
The radio technique is a promising method for detection of cosmic-ray air showers of energies around $100\,$PeV and higher with an array of radio antennas.
Since the amplitude of the radio signal can be measured absolutely and increases with the shower energy, radio measurements can be used to determine the air-shower energy on an absolute scale.
We show that calibrated measurements of radio detectors operated in coincidence with host experiments measuring air showers based on other techniques can be used for comparing the energy scales of these host experiments.
Using two approaches, first via direct amplitude measurements, and second via comparison of measurements with air shower simulations, we compare the energy scales of the air-shower experiments Tunka-133 and KASCADE-Grande, 
using their radio extensions, Tunka-Rex and LOPES, respectively.
Due to the consistent amplitude calibration for Tunka-Rex and LOPES achieved by using the same reference source, this comparison reaches an accuracy of approximately $10\,\%$ - limited by some shortcomings of LOPES, which was a prototype experiment for the digital radio technique for air showers.
In particular we show that the energy scales of cosmic-ray measurements by the independently calibrated experiments KASCADE-Grande and Tunka-133 are consistent with each other on this level.
\end{abstract}


\end{frontmatter}


\section{Introduction}
Cosmic rays are charged, high-energy particles from space
which offer a window to the most energetic processes in the universe.
Their origin remains uncertain, as they are deflected by magnetic fields 
during propagation and thus do not point back to their sources.
Instead, sources or source populations have to be identified indirectly by comparing the measured flux per energy and the mass composition to model predictions~\cite{AugerUpgradeICRC2015}.
The flux of cosmic-ray particles at high energies, above $10^{15}$\,eV, 
is too low for direct measurements, 
but instead has to be reconstructed from air showers, induced in the Earth's atmosphere, 
measured with extended devices on ground.
As cosmic-ray observables like the flux or mass composition are always interpreted as a function of energy, 
a precise and accurate energy measurement is of importance to all cosmic-ray detectors.

There are different methods for detecting air showers, 
of which most can be classified in particle detector arrays and optical techniques.
Particle detector arrays measuring the secondary particles at the observation level can be operated around-the-clock, 
and thus offer the highest exposure and best event statistics. 
But they are limited by systematic uncertainties from air-shower simulations based on hadronic interaction models beyond the energy range probed by accelerators, which a required for proper interpretation of the data.
Especially the muonic component of air showers seems to be poorly 
described by contemporary models~\cite{KASCADEGrandeMuons2015, AugerMuons2015}, 
possibly also distorting the energy scale of the detectors.
Optical techniques, detecting the air-Cherenkov or fluorescence light of the electromagnetic air-shower component
suffer less from systematic uncertainties of air-shower simulations,
but can only operate during clear and dark nights, reducing the statistics by an order of magnitude.
To overcome these problems, contemporary observatories combine advantages 
from the different observation techniques in hybrid detectors~\cite{AugerNIM2015, TelescopeArray2008}.

The radio detection technique is a promising detection method for high-energy air showers, 
which experienced a renewed burst of interest in the 2000s~\cite{SchroederReview2016,HuegeReview2016}.
Mainly due to geomagnetic deflection of the charged particles in the air shower, 
and to a lower extent also due to a time-varying charge excess in the shower front,
a radio signal in the MHz range is emitted~\cite{Kahn1966,Askaryan1962}.
Due to the special coherence conditions at the Cherenkov angle the emission by these mechanisms extents even up to the GHz range at this angle.
This has been confirmed by recent measurements indicating that the radio emission is beamed in the forward direction of the shower not only at MHz, but also GHz frequencies \cite{Smida:2014sia}.
Above 10$^{17}$\,eV the radio signal at MHz frequencies can be detected with an array of radio antennas.
With a full duty cycle and competitive precision, the radio detection technique 
combines advantages of various existing techniques.
Its low dependence on details of the atmospheric conditions and on the muon content 
of the air shower makes it particularly suitable for an accurate measurement of the shower energy,
which for a hybrid detector is already possible with a very sparse array~\cite{hiller2016singleantenna}.

The energy measurement via the radio signal is connected to its amplitude scale, i.e., 
the strength of the electric field emitted by the shower, as demonstrated by several 
experiments~\cite{Apel2014,Lautridou2013,Bezyazeekov2016,AERAenergyPRD2016}. 
In particular the possibility to measure the radio amplitude on an absolutely calibrated 
scale~\cite{Nehls2008, Bezyazeekov2015, NellesLOFARcalibration2015, AERAenergyPRL2016}, 
thus enables an absolute measurement of the shower energy.
As the present radio detectors are mainly operated together with host detectors,
also the energy scales of their hosts can be compared to each other via the calibrated radio measurements.
In this paper we present two methods to perform exemplarily such a comparison:
first via the energy estimator of the radio detectors and, second,  
via comparison to CoREAS simulations of the radio emission from air showers.
Using these methods, the energy scales of the non-imaging air-Cherenkov array Tunka-133~\cite{Berezhnev2012} and 
the particle-detector array KASCADE-Grande~\cite{Apel2010} are compared to each other, or more precisely the scale of the cosmic-ray energy spectra measured by these experiments around an energy of $10^{17}\,$eV of the primary particles. 
Tunka-133 and KASCADE-Grande are hosts to the radio extensions Tunka-Rex~\cite{Bezyazeekov2015} and LOPES~\cite{Falcke2005}, respectively.
For Tunka-Rex and LOPES the comparison is especially accurate and straight forward in interpretation,
because both experiments were calibrated with the same reference source~\cite{Apel2015}.
However, there are also some limitations, since LOPES was a prototype experiment in the noisy environment of a research center, since it covered only a small part of the KASCADE-Grande area, and since its antenna model has some shortcomings.
This analysis sheds light on the systematic effects originating from the independent energy calibrations of both experiments
and thus facilitates a combined interpretation of data from both experiments.

\section{Calibration}
\label{sec:calibration}
The absolute scales of the radio amplitude measured by Tunka-Rex and LOPES have been defined by a calibration with a reference source.
Descriptions of the exact process can be found in Refs.~\cite{Bezyazeekov2015} and 
\cite{Nehls2008,Apel2015}, respectively.
Both experiments used the same reference source.
Therefore, the dominating uncertainty of the calibration, the amplitude scale of the reference source itself, cancels out when comparing both experiments.
The remaining uncertainties of the amplitude scale that do not cancel out over multiple events and different antennas,
are $6\,\%$ from the temperature dependence of the reference source and 
$3\,\%$ from source positioning and alignment~\cite{Bezyazeekov2015}.
Moreover, there are uncertainties of several percent due to the dependence of the LOPES antenna gain on varying ground conditions~\cite{Nehls2008}, but the net effect is small, since the present analysis averages over many events recorded during different ground conditions.
Simulations are used to describe the angular dependence of the antenna response, which have shortcomings in the description of the zenith dependence of the LOPES antenna gain~\cite{SchroederICRC2015}. 
Unfortunately, a more accurate calibration of the zenith dependence is not possible, because LOPES is dismantled since 2012.
In summary, we estimate the uncertainty from the antenna calibration to $7\,\%$ for the amplitude scale of Tunka-Rex and LOPES relative to each other, with an additional uncertainty of the order of  $10\,\%$ from the zenith dependence of the LOPES antenna model.
The calibration uncertainty constitutes a correlated systematic uncertainty for the two methods of comparing the energy scales introduced in the next sections.

The energy calibrations of Tunka-133 and KASCADE-Grande are both performed 
with the help of air shower simulations.
The Tunka-133 calibration is based on the QUEST experiment which measured air-Cherenkov light of air showers coincidentally with the particle-detector array EAS-TOP~\cite{Korosteleva2007,Prosin2014}, which itself was calibrated with CORSIKA simulations using different hadronic interaction models, among them QGSJET~\cite{EASTOPenergySpectrum1999, QUEST_ICRC2003}.
KASCADE-Grande is calibrated with a newer version CORSIKA using different interaction models. 
For the purpose of this comparison the calibration based on QGSJET II is used, since this was used for previously published results~\cite{Apel2015, Apel2012}. 
Despite some known deficits \cite{AugerMuons2015, ArteageICRC2015}, QGSJET II is still one of the best hadronic models for air-shower simulations and widely used.

\section{Comparison of the energy scales via a radio energy estimator}
\label{sec:amplitude}
The energy scales of the host experiments can be compared directly via 
the measurement of the absolute amplitude scale of the radio signal in conjunction with shower energy reconstructed by the host experiment.
This can be done with any of the energy estimators typically used for radio detection of air showers, 
and for this analysis we have chosen the amplitude at a characteristic distance from the shower axis, 
since this has already been used by both, LOPES and Tunka-Rex, as energy estimator~\cite{Apel2014, Bezyazeekov2016}.
The LOPES experiment used the amplitude of the radio signal at a distance of $100\,$m.
This distance has been tuned to maximize the precision of the energy reconstruction for a typical event selection of LOPES. 
The amplitude at this distance features little dependence on the distance to the shower maximum, i.e., little dependence on the 
zenith angle and the atmospheric depth of the shower maximum of an air shower. 
Therefore, the amplitude at this distance is also a good choice for the comparison to another experiment like Tunka-Rex.

For LOPES the events used in this paper were acquired from the end of 2005 to the end of 2009, 
and triggered by the KASCADE particle detector array. 
Only events with an energy reconstructed by KASCADE-Grande above $10^{17}$\,eV, a zenith angle below \deg{40}, and a shower core inside the fiducial area of KASCADE are used (like in Ref.~\cite{Apel2014}). 
Additionally events disturbed by nearby thunderstorms are excluded~\cite{ApelLOPESthuderstorm2011}. 
The resulting events are analyzed with the standard analysis pipeline of LOPES applying certain quality cuts, 
e.g., requiring a minimum signal-to-noise ratio (see Ref.~\cite{ApelLOPES_CoREAS2013} for details), 
and 178 events pass all quality criteria.
To allow for sufficient event statistics in the radio-loud environment of the LOPES experiment, this signal-to-noise criterion is less strict than the one for Tunka-Rex, which implies larger per-event uncertainties for LOPES.  
The reconstructed signal of LOPES is limited to the effective band of $43$ to $74\,$MHz, 
and only the signal in the east-west aligned antenna was evaluated.
A simple exponential function was used to determine the radio amplitude at $100\,$m distance from the shower axis, 
since the average effect of more subtle features of the radio footprint (e.g., its east-west asymmetry and a bump at the Cherenkov angle) 
is only a few percent for this data set at this distance~\cite{ApelLOPES_CoREAS2013}.
Finally, the amplitude at $100\,$m was divided by the sine of the geomagnetic angle in order to normalize 
for the direction dependence of the strength of the geomagnetic radio emission.

This normalized amplitude is proportional to the energy of the air shower determined by the host experiment with a median signal amplitude per energy of 
$\dind{k}{100}{LOPES}=724\pm 12\,\mathrm{\frac{\upmu V/m}{EeV}}$.
The median is used here to reduce the impact of single outlier events of unknown origin already seen in earlier LOPES analyses.
To account for the difference in geomagnetic field between the LOPES and Tunka sites, we assume that the amplitude of the radio signal is proportional to the magnetic field strength, and divide $\dind{k}{100}{LOPES}$ by the value at the LOPES site of $47\,\upmu$T:
$\ind{\kappa}{LOPES}=15.40\pm0.26\,\mathrm{\frac{\upmu V/m}{{\upmu}T\ EeV}}$. 
The approximate proportionality of the radio amplitude with the geomagnetic Lorentz force has been confirmed by many experiments \cite{SchroederReview2016,HuegeReview2016}. 
Recently, slight deviations from an exactly proportional scaling with the magnetic field strength have been discussed based on simulations \cite{GlaserTheoryPaper2016}.
If true, this would change our result for the ratio $\ind{\kappa}{TRex}/\ind{\kappa}{LOPES}$ by about $2\,\%$, and consequently is negligible against other uncertainties.

A corresponding analysis has been performed for Tunka-Rex measurements with an event selection following the 
standard reconstruction method described in Ref.~\cite{Bezyazeekov2016}.
With the used selection criteria, both experiments have an energy threshold around 10$^{17}$\,eV. 
Though the radio detection is not fully efficient at this threshold, the triggering host detectors are. 
Because of the low duty cycle of the air-Cherenkov array Tunka-133, and the shorter run time of Tunka-Rex compared to LOPES, the available event statistics is similar for both experiments, although Tunka-Rex covers a much larger area.

For Tunka-Rex, the selection yields 196 Tunka-133 events from October 2012 until April 2014 with energies above 10$^{16.5}$\,eV, zenith angles $\theta\leq50^{\circ}$, and successful reconstruction of the radio signal.
This implies the application of standard quality cuts used in other Tunka-Rex analyses (see Refs.~\cite{Bezyazeekov2015,Bezyazeekov2016}), in particular a certain signal-to-noise ratio in at least three antennas and an agreement of the arrival directions reconstructed by the radio and the air-Cherenkov arrays.
As only difference to the standard Tunka-Rex pipeline the frequency range has been digitally limited to $43$ to $74\,$MHz after inverting the hardware response, i.e., the Tunka-Rex data have been evaluated inside the smaller effective band of LOPES, instead of the usual effective band of $35$ to $76\,$MHz of Tunka-Rex.

\begin{figure*}[tb]
  \includegraphics[width=0.5\textwidth]{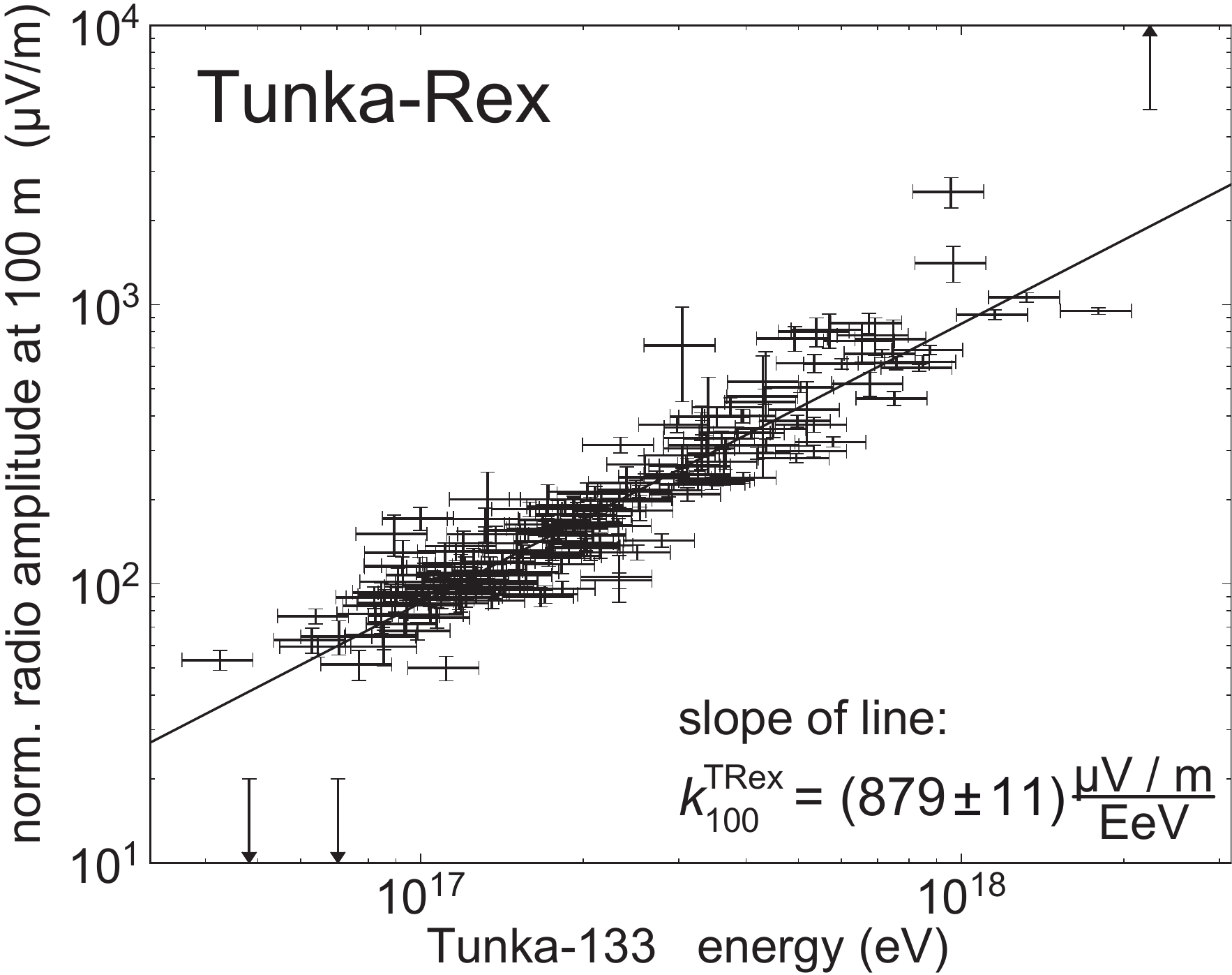}
  \includegraphics[width=0.5\textwidth]{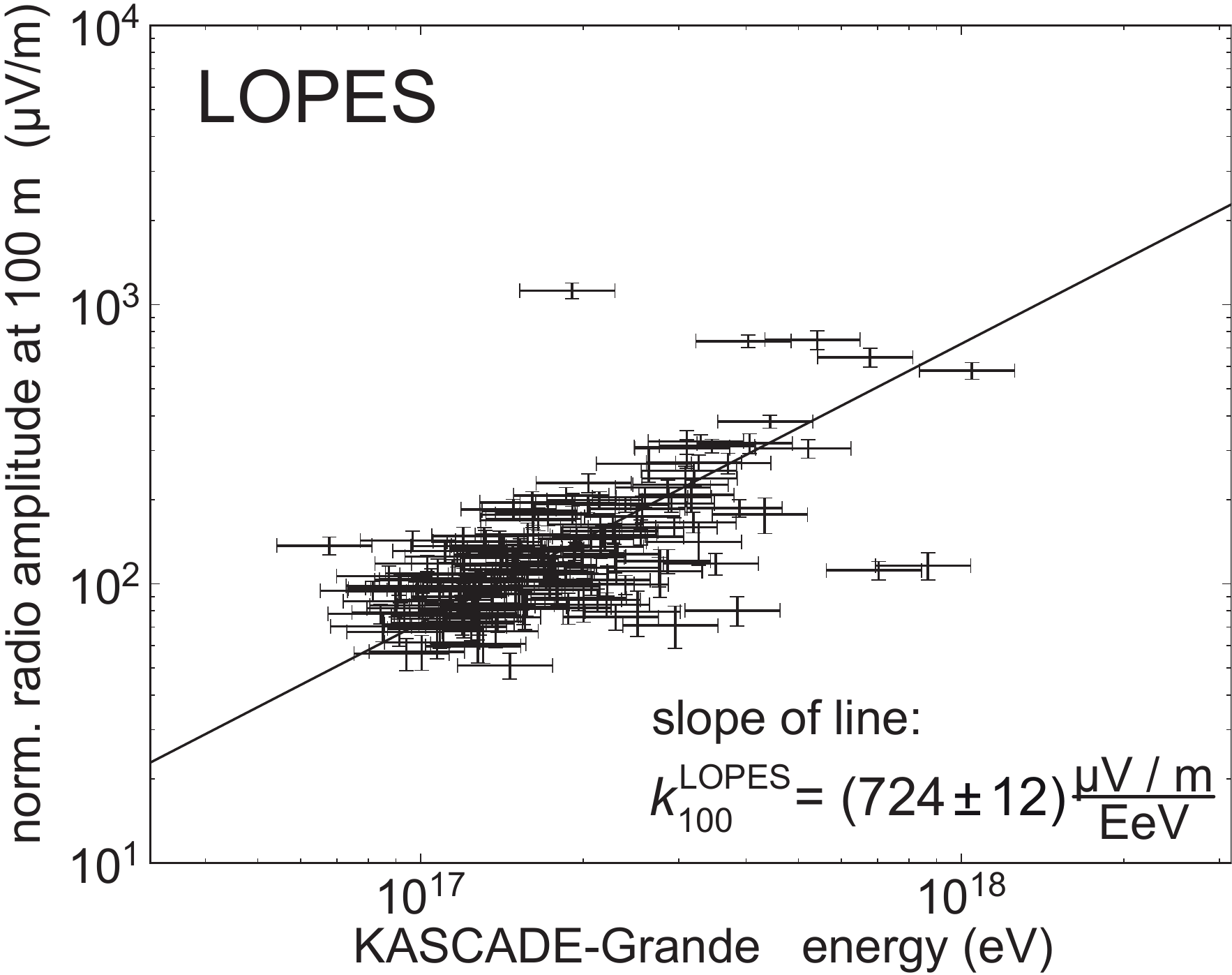}
  \caption{
  Amplitude at $100\,$m measured by the radio arrays Tunka-Rex (left) and LOPES (right) versus the shower energy reconstructed by their host experiments Tunka-133 and KASCADE-Grande, respectively, after division by the sine of the geomagnetic angle.
  The line indicates the median of the amplitude per energy, which is used to compare the amplitude scales to each other.
  }
  \label{fig:TRexvLOPES}
\end{figure*}

From the resulting event selection the reconstructed east-west component normalized to the sine of the geomagnetic angle was evaluated.
As for the LOPES analysis, the amplitude at $100\,$m distance from the shower axis has been determined using a simple exponential function for its lateral distribution.
The correction for the small azimuthal asymmetry of the footprint, usually applied for Tunka-Rex \cite{KostuninTheory2015}, was omitted here as it is also not applied for LOPES, and since it has been shown to have negligible impact on statistical analyses averaging over many events \cite{ApelLOPES_CoREAS2013}.
The resulting plots of amplitude versus the energy determined by the host experiments are shown in Fig.~\ref{fig:TRexvLOPES}.
The median of the obtained amplitude per energy is $\dind{k}{100}{TRex}=879\pm 11\,\mathrm{\frac{\upmu V/m}{EeV}}$ for Tunka-Rex which after normalizing to the magnetic field strength of 60\,$\upmu$T results in $\ind{\kappa}{TRex}=14.65\pm 0.18\,\mathrm{\frac{\upmu V/m}{{\upmu}T\ EeV}}$.

In order to compare the obtained values to each other, 
the difference in observation level between Tunka-Rex and LOPES has to be taken into account. 
The radio emission is generated mainly around the shower maximum, which in the observed energy range typically is at higher altitudes than the observation levels of both experiments. 
Thus, the main effect is that the radio emission is spread over a larger area for deeper observation levels reducing the amplitude at a given distance to the shower axis. 
This means that even at the chosen characteristic distance of $100\,$m the amplitude depends slightly on the distance from the detector to the shower maximum, which itself depends on the altitude of the detector and for each individual event on the atmospheric depth of the shower maximum and on the zenith angle. 
Since the effect of shower-to-shower fluctuations of the depth of shower maximum average out over the event statistics, only the zenith angle effect and the altitude of the detectors play a role here.

\begin{figure*}[tb]
\centering
  \includegraphics[width=0.7\textwidth]{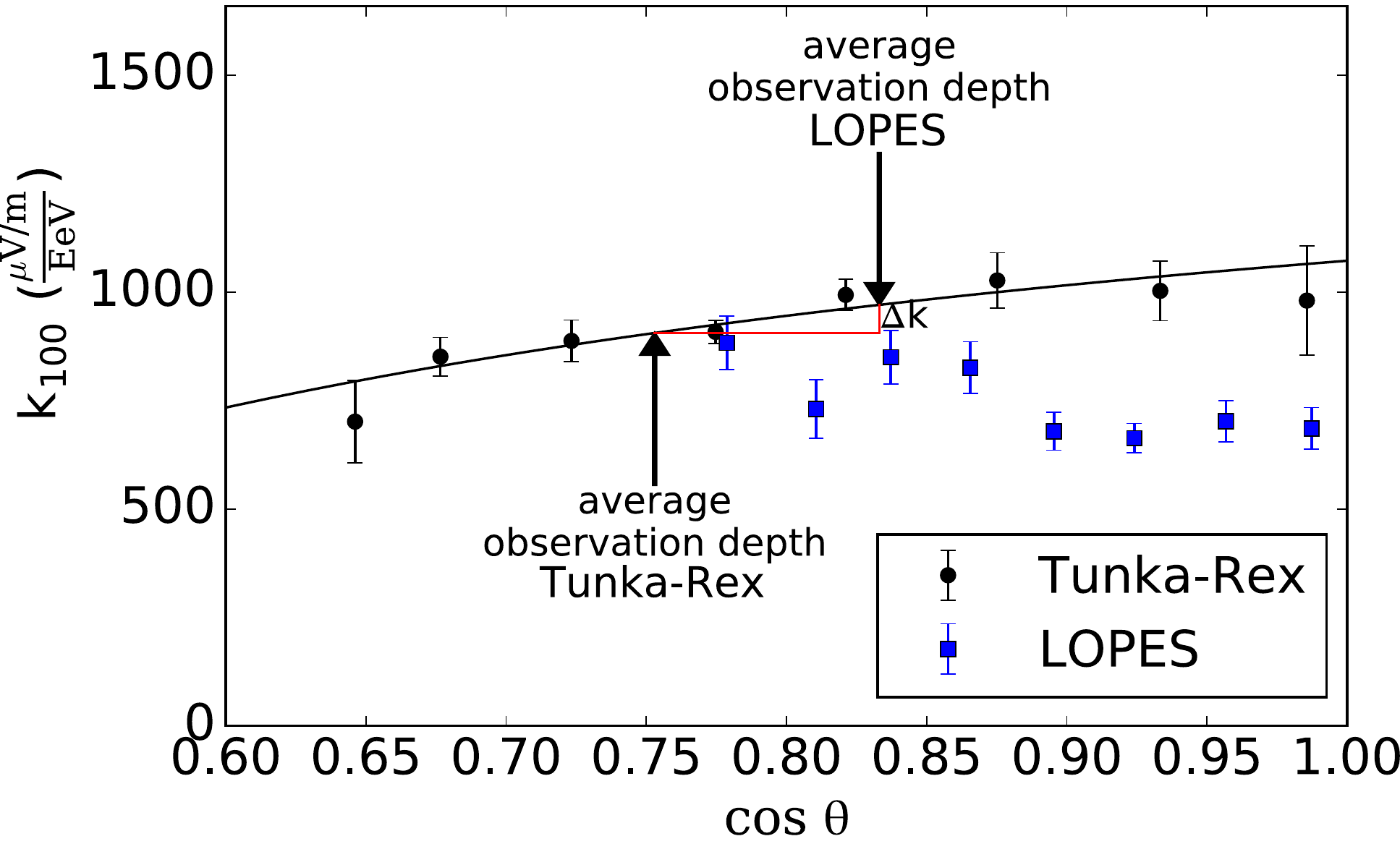}

  \caption{Amplitude per energy of the east-west component of Tunka-Rex and LOPES versus zenith angle.
  With the indicated zenith angles corresponding to the average observation depths, the systematic uncertainty due to the difference in observation depth is estimated. 
  The reason for the overall shift between the Tunka-Rex and LOPES data is the different geomagnetic field strength and energy scale of the host experiments.
  The reason for the different trend in the LOPES data likely is the deficient antenna model applied to the LOPES measurements.
  }
  \label{fig:amp_vs_theta}
\end{figure*}

The LOPES event distribution has an average zenith angle of \deg{27}. Due to the higher observation altitude of Tunka-Rex, air showers with \deg{27} zenith angle measured with Tunka-Rex would, however, have a smaller distance to the shower maximum than at LOPES and consequently a steeper footprint.
Instead, showers with \deg{35} zenith angle at Tunka-Rex have about the same distance to the shower maximum and are expected to have a similar radio signal on ground.
This angle is by chance close to the average zenith angle of the Tunka-Rex event distribution of \deg{41}.
The remaining systematic effect in the present analysis can be estimated by the average difference in amplitude at the characteristic distance of $100\,$m between \deg{35} and \deg{41} zenith angle. 
As shown in Fig.~\ref{fig:amp_vs_theta}, the resulting systematic uncertainty on the comparison of $\ind{\kappa}{TRex}$ and $\ind{\kappa}{LOPES}$ is approximately $7\,\%$, which is about the same as the systematic uncertainty from the calibration of the experiments.

In principle the effect can be corrected.
However, this makes only sense if all other systematic effects regarding the shower inclination are understood sufficiently well, e.g., a slight dependence of the energy scale of the host experiments on the shower inclination. 
In our case the dominating systematic uncertainty of about $10\,\%$ results from the deficient description of the zenith dependence of the LOPES antenna gain, which likely is why the LOPES data show a different trend over zenith angle.
Consequently, we take the size of the effect observed in the Tunka data as a method-specific systematic uncertainty of $7\,\%$, and additionally, $10\,\%$ uncertainty of the LOPES antenna model for all methods in the interpretation of our results in the Conclusion, Sec.~\ref{sec:conclusion}.

Since the measured energy $\ind{E}{m}$ from either experiment may have a systematic shift compared to the real energy $\ind{E}{real}$, 
the measured coefficient $\ind{\kappa}{m}$ deviates from the real one $\ind{\kappa}{real}$
\begin{equation}
\ind{\kappa}{m}=\frac{\ind{E}{real}}{\ind{E}{m}}\cdot \ind{\kappa}{real}
\end{equation}   
Thus, the energy scales of Tunka-Rex and LOPES and their hosts, KASCADE-Grande and Tunka-133, 
can be compared to each other using the radio measurements of $\ind{\kappa}{m}$
\begin{equation}
f_{\mathrm{amp}}=\frac{\ind{E}{KG}}{\ind{E}{T133}}=\frac{\ind{\kappa}{TRex}}{\ind{\kappa}{LOPES}}.
\end{equation}
The resulting ratio of reconstructed energies is $\ind{f}{amp}=0.95\pm0.07$ for this method of comparing radio amplitudes, i.e, the energy scale of KASCADE-Grande is ${(5 \pm 7)\,\%}$ lower than the energy scale of Tunka-133.
This uncertainty includes only method-specific contributions, which are dominated by the systematic effect due to the difference in observation depth.

\section{Comparison of the energy scales via CoREAS simulations}
\label{sec:simulations}

\begin{figure}[t]
 \centering
  \includegraphics[width=0.49\textwidth]{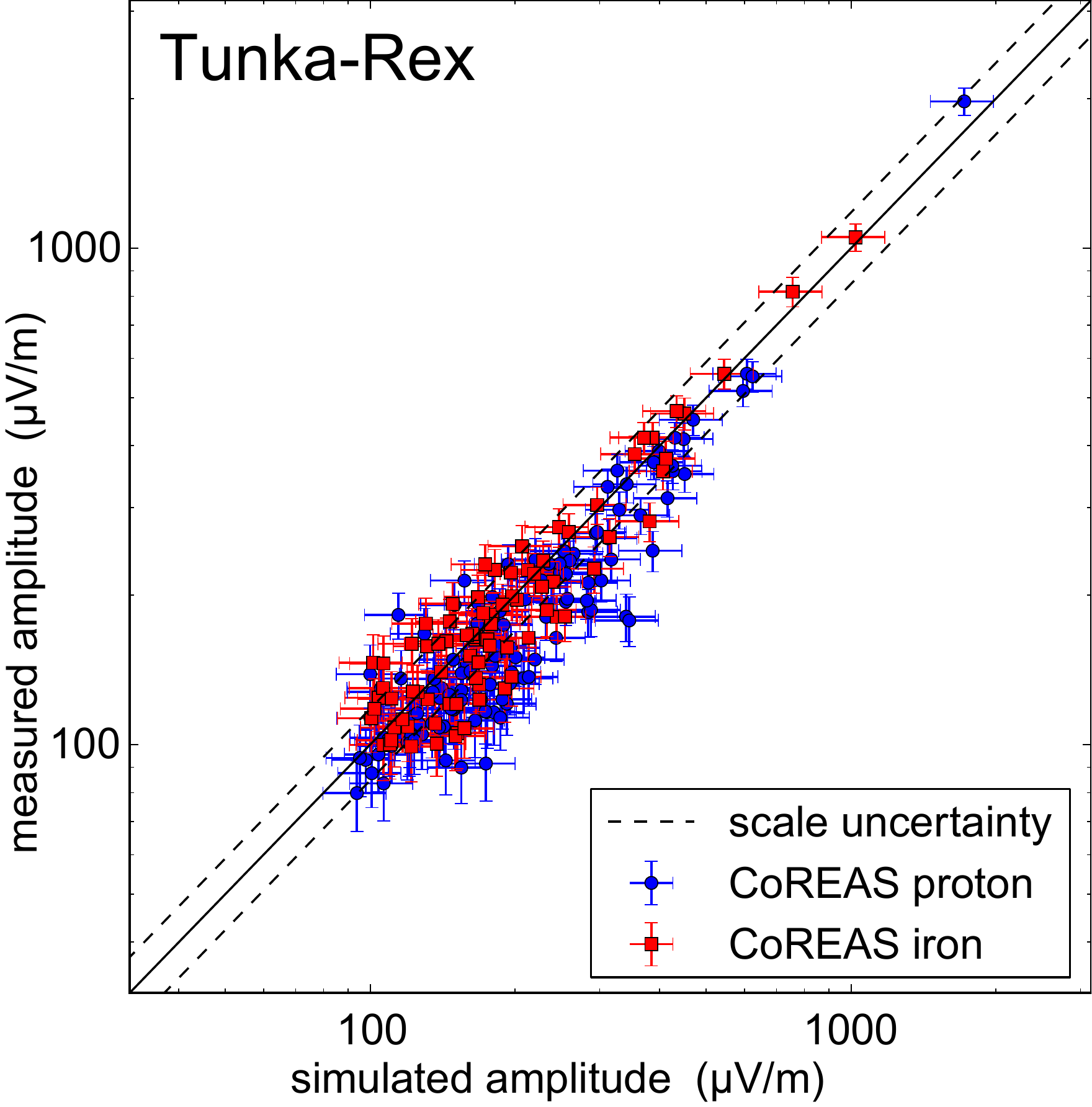}
  \hfill
  \includegraphics[width=0.49\textwidth]{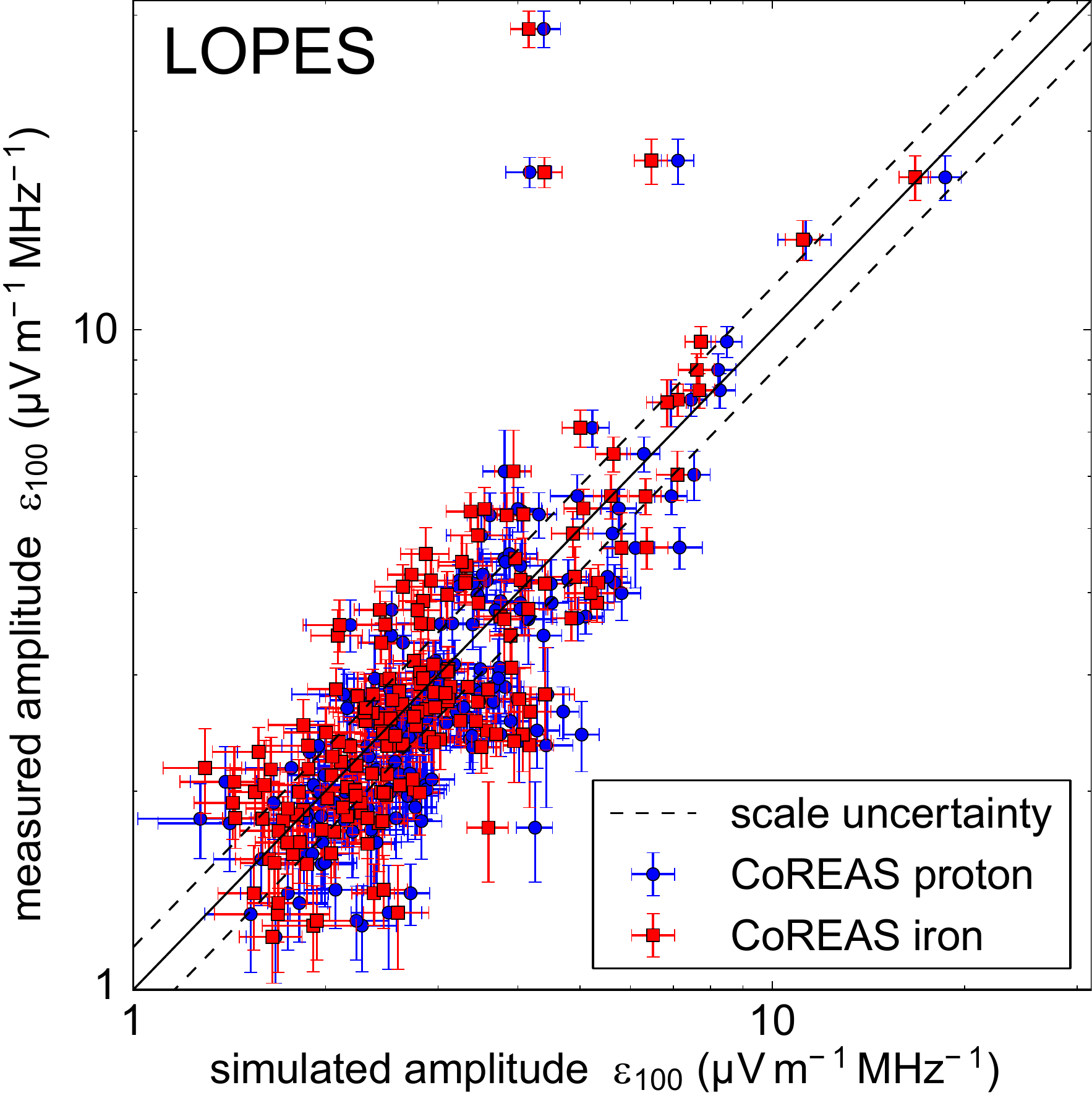}
  \caption{Radio amplitude at each Tunka-Rex station with signal (left), and at $100\,$m distance from the shower axis for events measured by LOPES (right) over the amplitudes simulated by CoREAS (the left figure is slightly modified from Ref.~\cite{Bezyazeekov2015}). 
  For both experiments the standard reconstruction pipelines are used, i.e, Tunka-Rex amplitudes are the maximum absolute values of the electric-field vector in the band of $35-76\,$MHz, and LOPES amplitudes are the maximum of a Hilbert envelope to the east-west component in the band of $43-74\,$MHz.)
  }
  \label{fig:LOPESvsCoREAS}
\end{figure}    

\begin{figure*}[tb]
  \includegraphics[width=0.5\textwidth]{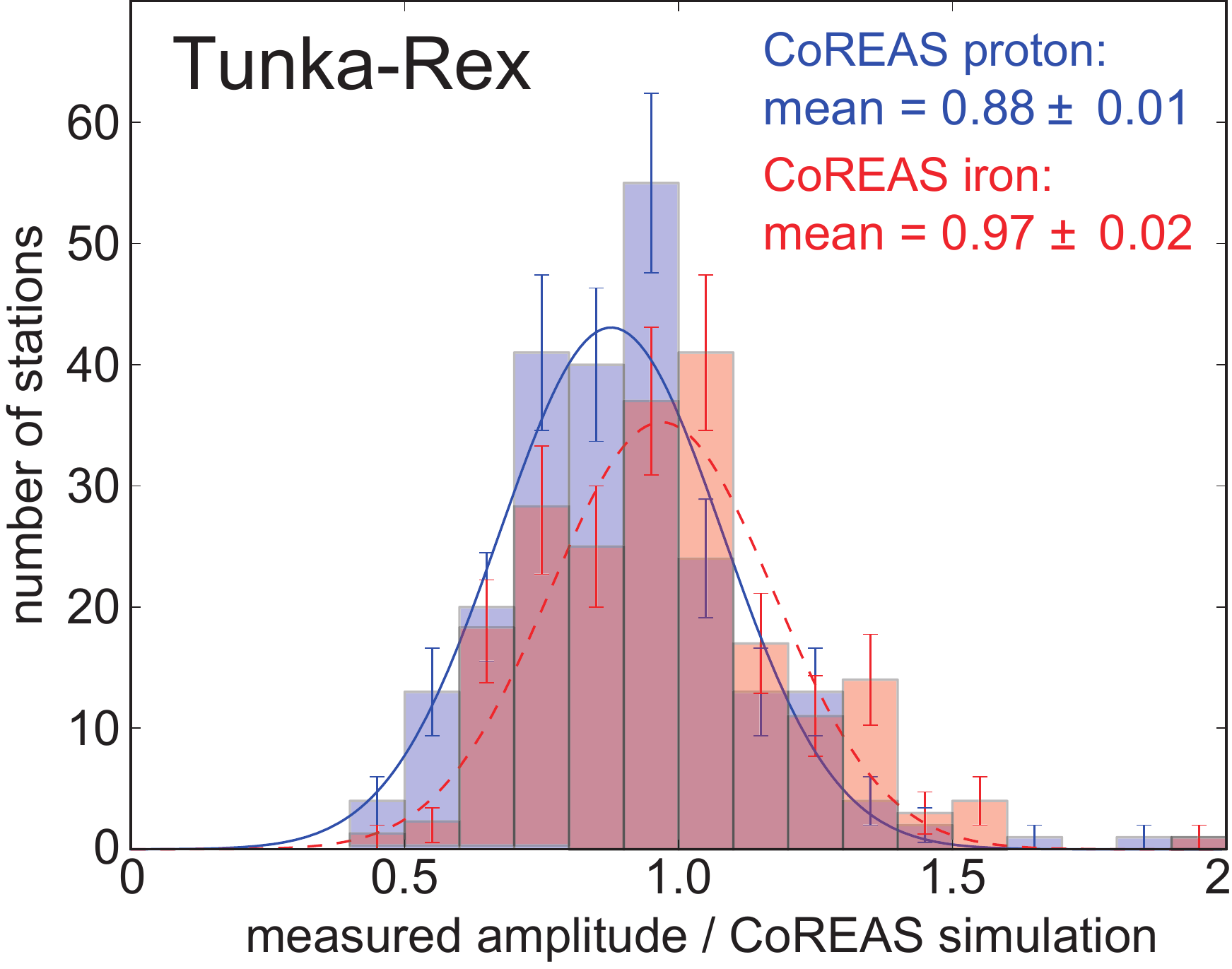}
  \includegraphics[width=0.5\textwidth]{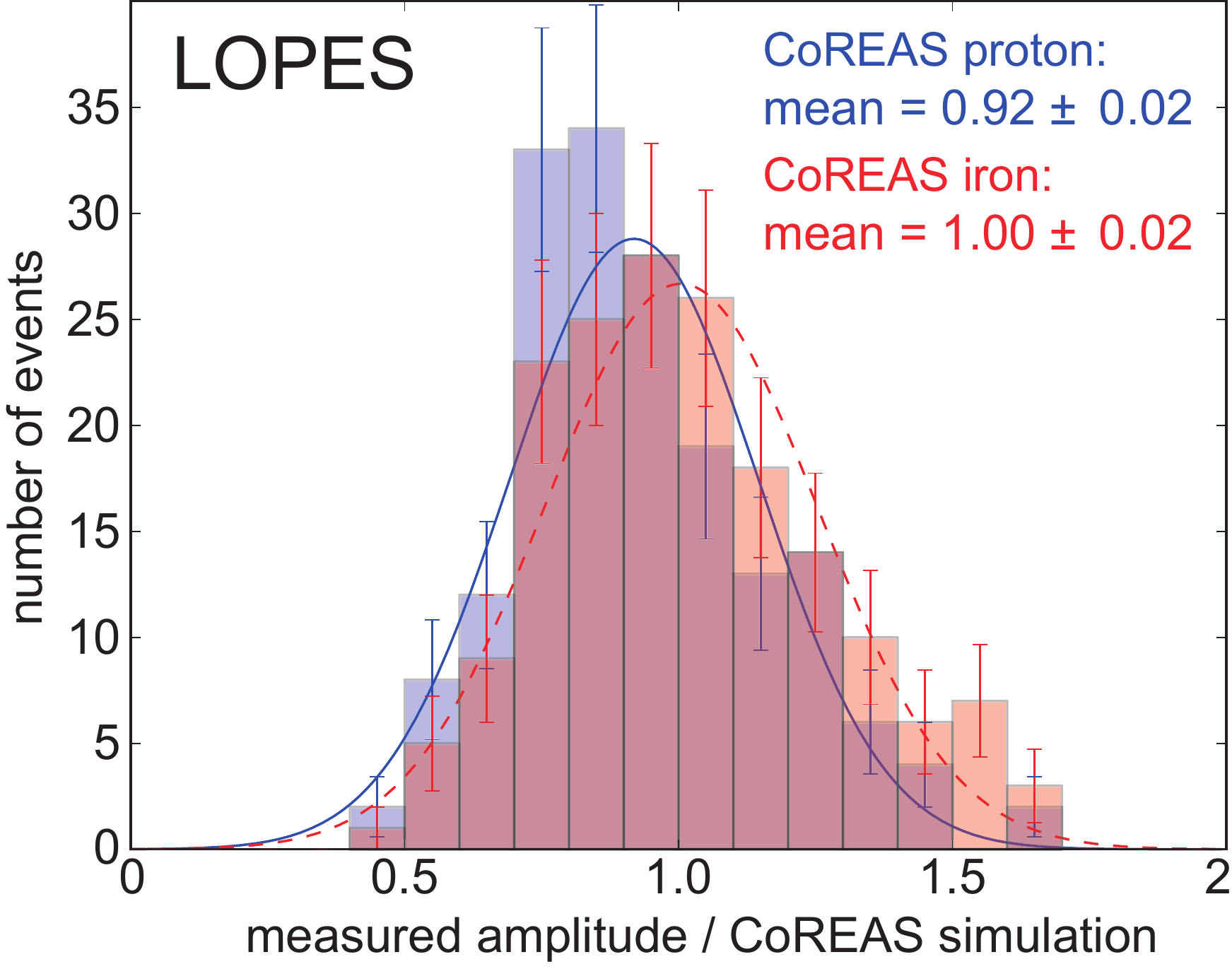}
  \caption{
  The ratio of reconstructed amplitudes of Tunka-Rex (left) and LOPES (right) versus predictions from air-shower simulations with CoREAS for protons and iron nuclei as primary particles.
  For Tunka-Rex the depth of the shower maximum is tuned in the simulations to the measured one and the amplitude at each station is compared.
  For LOPES instead the amplitude at $100\,$m is compared for each event, which is relatively independent of the depth of the shower maximum.
  The mean of a Gaussian distribution obtained from a fit is used to define the energy scale using CoREAS as a benchmark.
  }
  \label{fig:TRexvCoREAS}
\end{figure*}       
Another way to compare LOPES to Tunka-Rex is by using simulations of the radio emission from air showers as a benchmark, as long as the same simulation code is used.
By configuring the simulations according to the respective site the difference in magnetic field and observation depth is automatically taken into account in this case.
For the present analysis we used the CoREAS~\cite{CoREAS} radio extension of the CORSIKA simulation program for air showers and the hadronic interaction model QGSJET in the versions II.03 for LOPES and II.04 for Tunka-Rex, where both versions have almost negligible difference for the radio emission.
Since the type of primary particle a priori is unknown, for each measured event two simulations have been performed, one each for the extreme case of a proton and an iron nucleus as primary particle. 
The energy and arrival direction are set to the values reconstructed by the host experiments KASCADE-Grande and Tunka-133, respectively, i.e., the energy scale of the host experiments is input to the simulations.

For LOPES we used the CoREAS simulations already presented in Refs.~\cite{ApelLOPES_CoREAS2013, Apel2015}, 
and applied two additional improvements compared to these references, which make the comparison slightly more accurate (the effect is only a few percent and unimportant for the previous analyses compared to their systematic uncertainties). 
First, since the energy calibration of KASCADE-Grande was slightly improved over time, the simulated amplitudes were rescaled linearly according to the energy shift between the KASCADE-Grande calibration used for production of the simulations and the one of Ref.~\cite{Apel2012} used in this paper. 
Second, we now applied a full detector simulation to the CoREAS output, i.e., the simulated amplitudes are directly comparable to the measured ones~\cite{LinkICRC2015}. 
Unfortunately, KASCADE-Grande features no measurement of the depth of shower maximum, and the simulations likely have a different depth of shower maximum than the measured events, since this varies from shower to shower.
This is important because the distance to the shower maximum affects the slope of the radio lateral distribution~\cite{LOPES_MTD_Correlation2011}.
To minimize the impact only the amplitude at $100\,$m distance from the shower axis is used for the comparison, because at this distance the amplitude depends least on the depth of shower maximum, which is one of the reasons why the same distance has been selected for the energy estimator in the previous section.
The mean ratio between measured and simulated amplitudes at $100\,$m obtained with LOPES is
$\dind{F}{LOPES}{p}=0.92\pm0.02$ for proton and $\dind{F}{LOPES}{Fe}=1.00\pm0.02$ 
for iron primaries (see Fig.~\ref{fig:LOPESvsCoREAS}).

For Tunka-Rex we use the comparison of measurements from October 2012 to April 2013 with CoREAS simulations already shown in Ref.~\cite{Bezyazeekov2015}.
Since the simulations take into account the situation and response of the detectors, in contrast to the previous section the Tunka-Rex standard analysis can be and is used.
Another slight advantage of Tunka-Rex is that Tunka-133 provides a measurement of the depth of shower maximum, which has been used to select simulations whose depth of shower maxima are consistent within $30\,$g/cm$^2$ to the measured one.
Thus, the measured and simulated amplitudes can be compared for each antenna individually irrespective of the distance from the shower axis.
As for LOPES, the simulated events undergo a full detector simulation, including antenna and hardware response,
downsampling and digitization, before adding measured noise and applying the same reconstruction algorithms as for the measured events.
The mean ratio between the amplitudes measured by Tunka-Rex and simulated by CoREAS is $\dind{F}{TRex}{p}=0.88\pm0.01$ for proton 
and $\dind{F}{TRex}{Fe}=0.97\pm0.02$ for iron primaries (cf. Fig.~\ref{fig:TRexvCoREAS}).

How is this connected to the energy scale? 
A systematic shift in the energy scale of the host experiments, which is used as input for the simulations, also shifts the ratio between measured and simulated amplitude by:
\begin{equation}
F=\frac{\ind{E}{real}}{\ind{E}{m}}\cdot \ind{F}{real}
\end{equation}
with ${\ind{E}{real}}$ the real energy in nature, ${\ind{E}{m}}$ the energy measured with the energy scale of the host experiment, and $\ind{F}{real}$ the ratio between the amplitude predicted by CoREAS and the real amplitude in nature.
Because Tunka-Rex and LOPES are compared to the same version of CoREAS, a possible constant scale mismatch between CoREAS and nature, 
$\ind{F}{real}$, cancels out when comparing both experiments with each other.
Thus, the derived ratio of energy scales is   
\begin{equation}
f_{\mathrm{sim}}=\frac{\ind{E}{KG}}{\ind{E}{T133}}=\frac{\ind{F}{TRex}}{\ind{F}{LOPES}}.
\end{equation}
The obtained ratios of scales are $\dind{f}{sim}{p}=0.96\pm0.05 $ and 
$\dind{f}{sim}{Fe}=0.97\pm0.06$ for the proton and iron simulations, respectively.
This means that the KASCADE-Grande energy scale is lower than the Tunka-133 energy scale by $(4 \pm 5)\,\%$ or $(3 \pm 6)\,\%$, respectively.
The uncertainties include the statistical uncertainty of around $2\,\%$ for each ratio and $5\,\%$ from the analysis method and the different versions of the hadronic interaction model, which are added in quadrature.
The uncertainties of the method arise because the ratio $F_{\mathrm{TRex}}$ varies by several percent, depending on details of the analysis procedure, 
such as bandwidth and model of the lateral distribution, which were not matched between both experiments for this analysis.


\section{Conclusion}
\label{sec:conclusion}
We have shown that energy scales of different air-shower experiments 
can be independently checked against each other by using accurately calibrated radio detectors. 
In particular we applied two different methods for this cross-check on the radio extensions LOPES and Tunka-Rex of the KASCADE-Grande and Tunka-133 air-shower arrays: 
one method which relies purely on measured data, but features a systematic uncertainty caused by the different observation levels, and another method based on simulations taking into account the differences between the experimental settings. 
In addition to the method-dependent uncertainties between $5\,\%$ and $7\,\%$ both methods share a correlated systematic uncertainty of $7\,\%$ due to the relative calibration of LOPES and Tunka-Rex. 
Finally, the insufficient description of the zenith dependence of the LOPES antenna gain constitutes a dominating systematic uncertainty of about $10\,\%$.
This shows the importance of accurate antenna calibrations for current and future experiments.
As combined result of both methods we show that the energy scales of KASCADE-Grande and Tunka-133 obtained by secondary-particle and air-Cherenkov-light detection, respectively, are consistent to an accuracy of the order of $10\,\%$ - limited by systematic uncertainties of the LOPES experiment.

\begin{figure}[tb]
  \centering
  \includegraphics[width=0.7\textwidth]{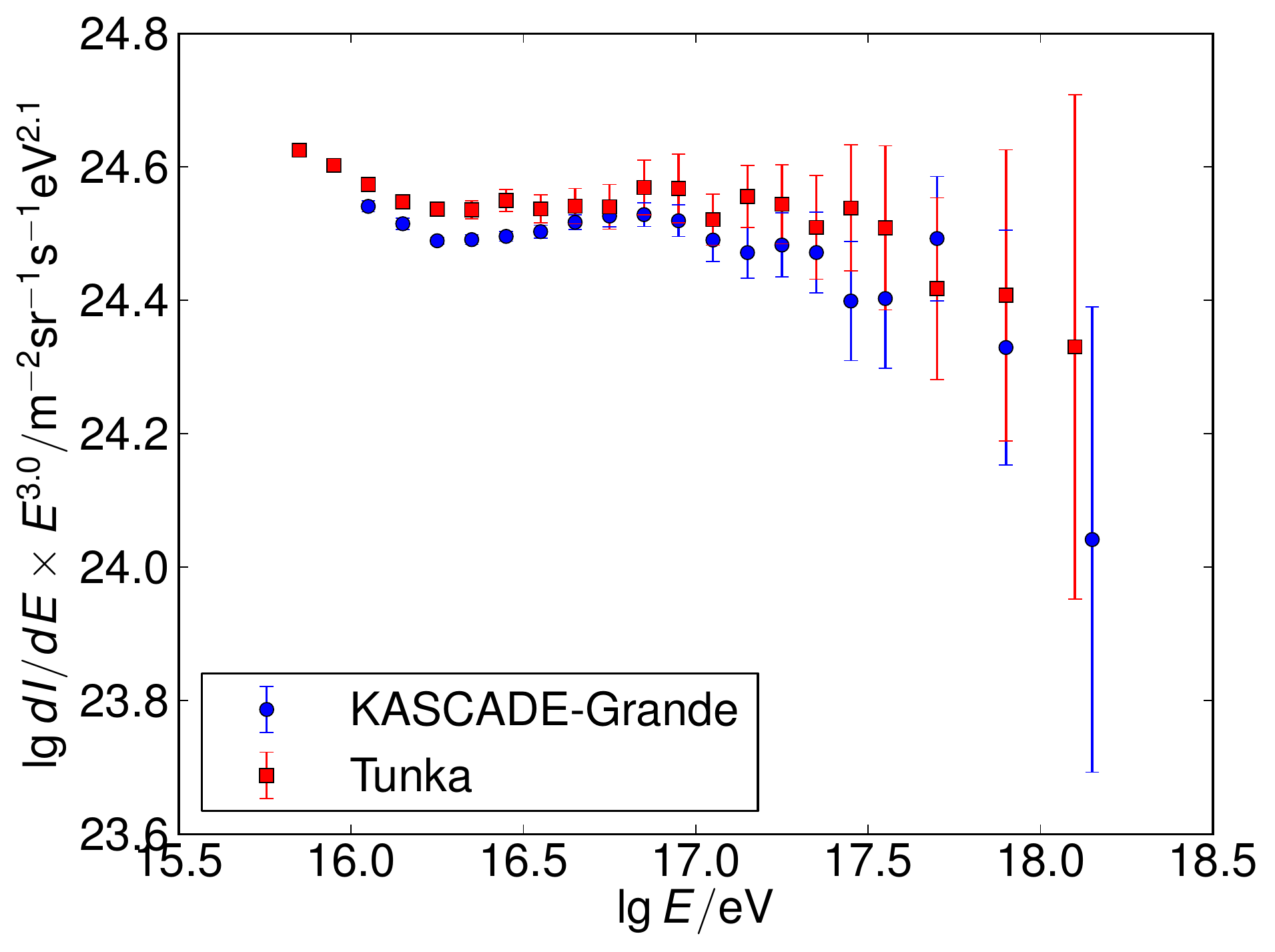}
  \caption{Energy spectra of cosmic rays from KASCADE-Grande~\cite{Apel2012} and Tunka-133~\cite{Prosin2014}: normalized flux per energy.
  The energy range also observed by the radio extensions Tunka-Rex and LOPES is $10^{17}$ to $10^{18}$\,eV.
  With a systematic increase of KASCADE-Grande energies by $4\,\%$ (or a corresponding decrease of Tunka-133 energies) the average flux per energy of both experiments can be brought to agreement in this energy range. 
  }
  \label{fig:GrandeTunka}
\end{figure}
\begin{figure}[tb]
  \centering
  \includegraphics[width=0.7\textwidth]{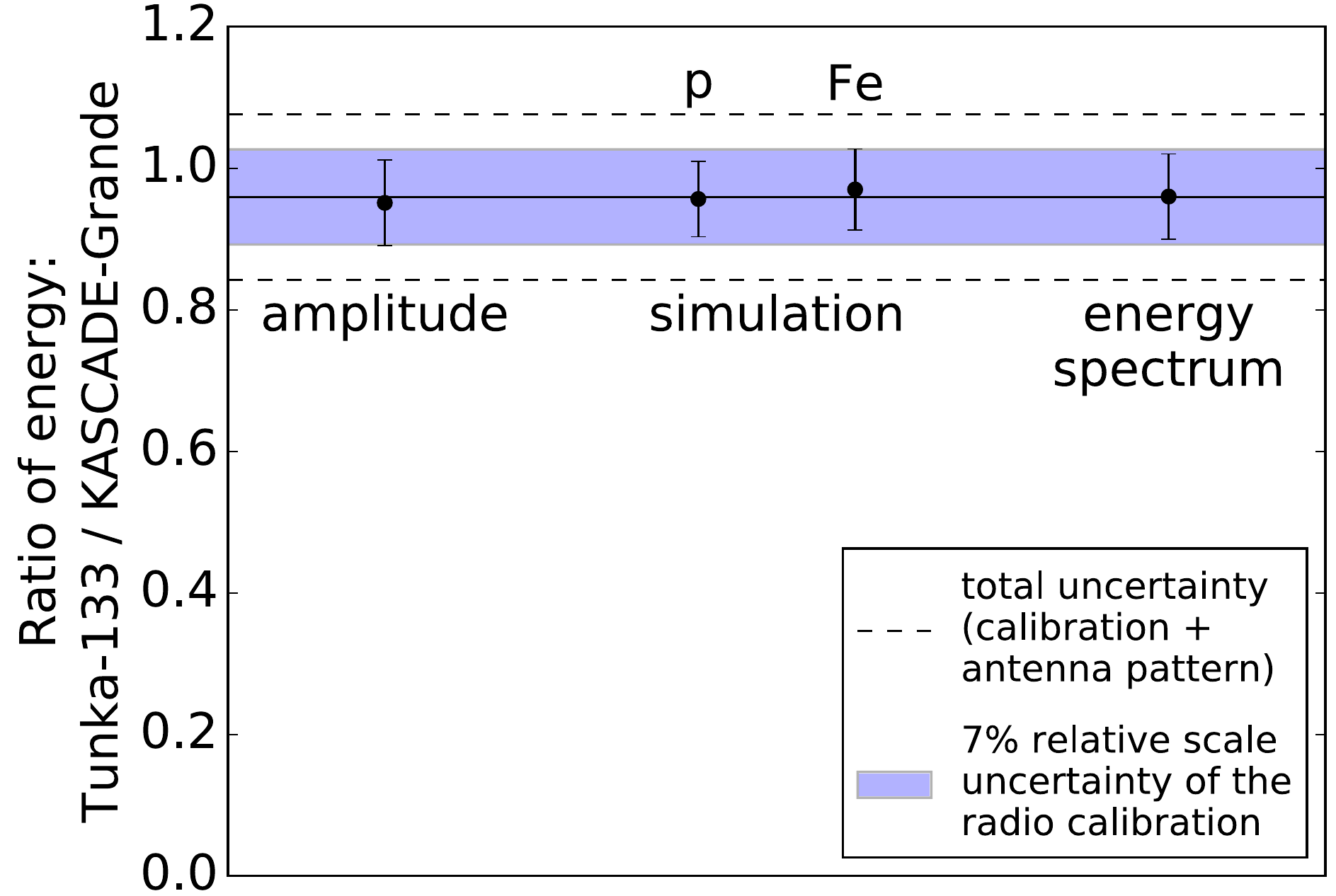}
  \caption{Results from the comparison of energy scales  
  between the experiments Tunka-Rex and LOPES, and their hosts Tunka-133 and KASCADE-Grande, respectively.
  The values \lq amplitude\rq, \lq simulation\rq, and \lq spectrum\rq~refer to the results presented in Secs.~\ref{sec:amplitude}, \ref{sec:simulations}, and \ref{sec:conclusion}, respectively, and the indicated uncertainties are discussed in Sec.~\ref{sec:calibration} and Sec.~\ref{sec:amplitude}
  }
  \label{fig:methods}
\end{figure}

To cross-check this claim, published energy spectra of KASCADE-Grande~\cite{Apel2012}
and Tunka-133~\cite{Prosin2014} are compared in the energy range of $10^{16.8}$ to $10^{18.0}$\,eV (see Fig.~\ref{fig:GrandeTunka}).
Assuming a simple, constant energy shift between both experiments, 
and given that both experiments measure the same cosmic-ray spectrum,
the spectra can be brought to match by shifting the KASCADE-Grande energy upwards by $4\,\%$ (or vice-versa down-shifting Tunka-133),
i.e., $\ind{f}{spec}=0.96\pm0.06$.
The deviation is not statistically significant and confirms the result obtained by
the radio measurements:
the energy scales of KASCADE-Grande and Tunka-133 are consistent and differ at most by about $10\,\%$, despite the fact that they have been obtained using two different measurement techniques, namely arrays of particle and air-Cherenkov detectors, respectively. 
Since both experiments rely on hadronic interaction models for the interpretation of their data, this also indicates that these interpretations are consistent.
The obtained results are summarized in Fig.~\ref{fig:methods}.

One astrophysical implication of this result is that the comparison of
features observed in the energy spectrum is now possible with smaller uncertainty, 
e.g., whether the knee in the heavy component of the energy spectrum observed by
KASCADE-Grande at about $10^{17}\,$eV~\cite{KGheavyKnee2011}, and a structure named as `second knee' observed 
by several experiments at about $3\cdot10^{17}\,$eV~\cite{HoerandelErice2006, Tunka133_UHECR2014} are one and the same or different features.

In the future, the accuracy of the presented methods can be further increased, e.g., by studying the systematic effects regarding the shower inclination and the observation levels of the experiments in more detail or by using different observables of the radio signal such as the integrated radiation energy~\cite{AERAenergyPRL2016}.
While this study assumes a constant value for the energy offset between the two experiments, given sufficient statistics and accuracy, the offset can in principle also be studied as a function of energy, or studied separately for different mass groups of the primary cosmic rays.

Moreover, the method can be easily applied to other air-shower arrays featuring a radio extension, in particular, AERA~\cite{AERAantennaPaper2012} at the Pierre Auger Observatory~\cite{AugerNIM2015}, and LOFAR~\cite{NellesLOFARcalibration2015}.
When further improving the calibration accuracy of the antenna arrays, radio measurements could also be used to calibrate air-shower detectors or to combine and compare data from different experiments on a common energy scale.
    
\section*{Acknowledgments}
The construction of Tunka-Rex was funded by the German Helmholtz association and the Russian Foundation for Basic Research (grant HRJRG-303). 
Moreover, this work was supported by the Helmholtz Alliance for Astroparticle Physics (HAP), by Deutsche Forschungsgemeinschaft (DFG) grant SCHR 1480/1-1, and by the Russian grant RSF 15-12-20022.
LOPES and KASCADE-Grande have been supported by the German Federal Ministry of Education and Research. 
KASCADE-Grande is partly supported by the MIUR and INAF of Italy, the Polish Ministry of Science and Higher Education and by the Romanian Authority for Scientific Research UEFISCDI (PNII-IDEI grant 271/2011). 
The authors acknowledge stimulating discussions within the \lq radio community\rq, in particular with colleagues of the Pierre Auger Observatory and of LOFAR.
    
\bibliography{scalerefs}

\end{document}